\documentclass{vgtc}                          
\ifpdf
  \pdfoutput=1\relax                   
  \usepackage{graphicx}                
  \DeclareGraphicsExtensions{.pdf,.png,.jpg,.jpeg} 
\else
  \usepackage{graphicx}                
  \DeclareGraphicsExtensions{.eps}     
\fi%

\graphicspath{{figures/}{pictures/}{images/}{./}} 

\usepackage{microtype}                 
\PassOptionsToPackage{warn}{textcomp}  
\usepackage{textcomp}                  
\usepackage{mathptmx}                  
\usepackage{times}                     
\usepackage{cite}                      
\usepackage{tabu}                      
\usepackage{booktabs}                  

\usepackage{mathtools}

\usepackage{enumitem}
\newlist{steps}{enumerate}{1}
\setlist[steps, 1]{label = Step \arabic*}
\usepackage{amsmath}
\usepackage{amssymb}
\usepackage{makecell}
\usepackage{multirow}
\usepackage{xcolor}
\usepackage{mathrsfs}
\usepackage[switch]{lineno}

\onlineid{0}

\vgtccategory{Research}

\vgtcinsertpkg

\title{Heter-Sim: Heterogeneous multi-agent systems simulation by interactive data-driven optimization}

\author{Jiaping Ren\thanks{e-mail: ren\_jia\_ping@zju.edu.cn}\\ %
        \parbox{2.4in}{\scriptsize \centering State Key Lab of CAD\&CG, Zhejiang University\\} %
\and Wei Xiang\thanks{e-mail: xiang\_wei@zju.edu.cn}\\ %
     \parbox{2.4in}{\scriptsize \centering State Key Lab of CAD\&CG, Zhejiang University}
\and Yangxi Xiao\thanks{e-mail: x\_x@zju.edu.cn}\\ %
     \parbox{2.4in}{\scriptsize \centering State Key Lab of CAD\&CG, Zhejiang University}
\and Ruigang Yang\thanks{e-mail: yangruigang@baidu.com}\\ %
     \parbox{2.4in}{\scriptsize \centering Baidu}
\and Dinesh Manocha\thanks{e-mail: dmanocha@gmail.com}\\
    \parbox{2.4in}{\scriptsize \centering Department of Computer Science and Electrical and Computer Engineering, University of Maryland}
 \and Xiaogang Jin\thanks{e-mail: jin@cad.zju.edu.cn}\\ %
     \parbox{2.4in}{\scriptsize \centering State Key Lab of CAD\&CG, Zhejiang University}}

\abstract{
Interactive multi-agent simulation algorithms are used to compute the trajectories and behaviors of different entities in virtual reality scenarios. However, current methods involve considerable parameter tweaking to generate plausible behaviors. We introduce a novel approach (Heter-Sim) that combines physics-based simulation methods with data-driven techniques using an optimization-based formulation.
Our approach is general and can simulate heterogeneous agents corresponding to human crowds, traffic, vehicles, or combinations of different agents with varying dynamics. We estimate motion states from real-world datasets that include information about position, velocity, and control direction. Our optimization algorithm considers several constraints, including velocity continuity, collision avoidance, attraction, and direction control. To accelerate the computations, we reduce the search space for both collision avoidance and optimal solution computation. Heter-Sim can simulate tens or hundreds of agents at interactive rates and we compare its accuracy with real-world datasets and prior algorithms. We also perform user studies that evaluate the plausible behaviors generated by our algorithm and a user study that evaluates the plausibility of our algorithm via VR.
}

\teaser{
  \centering
  \includegraphics[width=6.5in]{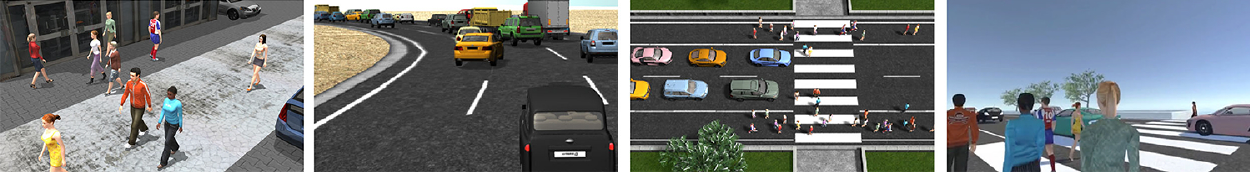}
  \caption{Our heterogeneous multi-agent simulation algorithm can be used for scenarios with tens or hundreds of different types of agents sharing a physical space. Pedestrians walking on a street (the first). Cars moving on a twisting road (the second). Traffic including cars and pedestrians (the third). Traffic shown through VR (the fourth). Our approach can generate plausible behaviors at interactive rates on a desktop PC and through VR.}
\label{fig:resultsAll}
}

\CCScatlist{
  \CCScatTwelve{Computing methodologies}{Computer graphics}{Animation}{Procedural animation};
  \CCScatTwelve{Computing methodologies}{Computer graphics}{Animation}{Physical animation};
}

\begin{document}

\firstsection{Introduction}

\maketitle

Many virtual reality and training systems need to be able to simulate different types of agents, including human crowds and traffic. Applications include VR therapy for crowd phobias, traffic agents for autonomous driving, urban design and planning, driving simulators for education and entertainment, etc. It is important to simulate the behaviors and trajectories of different types of agents, including pedestrians and vehicles, and the interactions between such heterogeneous agents. Furthermore, it is important to develop general plausible algorithms that are applicable to a wide variety of scenarios.

There is extensive work on interactive multi-agent simulation, including crowd simulation and traffic simulation. These works include techniques based on rule-based methods~\cite{Boids87Flocks}, physics-based simulations~\cite{wang2015bswarm,Karamouzas2014Universal}, energy-based models~\cite{dutra2017gradient}, data-driven methods~\cite{pettre2009Experiment,chao2018realistic}, and combinations of these approaches~\cite{sewall2011interactive,kim2016interactive}. These methods are flexible and have been successfully applied to different scenarios. However, they often use many parameters and require a significant amount of effort to achieve good results that are plausible and match the behaviors observed in real-world scenarios. Furthermore, the results of these methods often seem too regular because all the agents have similar locomotion or movement patterns.

With the improvement of data acquisition techniques, more data-driven methods are emerging. Most of these methods are patch-based or use real-world agent trajectories~\cite{jordao2014crowd,ju2010morphable,kim2016interactive,wang2015bswarm}. These methods extract patches or trajectory segments from input datasets and either connect them with some rules or use them to learn some characteristics of an agent's motion. With these methods, users can generate more plausible or more accurate results than with traditional rule-based or physics-based simulation methods. However, the variety of the simulation results depends on the amount of input data. If the amount of input data is small, the simulation results will be periodic and monotonous.

Most of the existing methods only apply to one kind of agent, e.g., only human pedestrians or only vehicles. In contrast, we want to use a general method to model the behaviors of different kinds of agents in a heterogeneous setting while retaining the motion features of each kind of agent. This is important in many situations like simulating the motion trajectories and interactions between cars and humans at a traffic crossing. Data-driven methods can also help us with simulating interactions between heterogeneous agents. However, data-driven methods depend on the input data, and it is difficult to simulate behavior in a scenario that is different from the one that generated the input data.

\noindent {\bf Main Results:} We present a novel, heterogeneous multi-agent simulation algorithm (Heter-Sim) that combines the benefits of prior data-driven and physics-based simulation methods to generate general and plausible simulations. Our interactive approach can simulate not only different kinds of agents while generating plausible behaviors, but also scenarios beyond those included in the input datasets. We convert various datasets captured using different types of sensors into a uniform format and extract the agents' states, including velocity information. We model the decision-making or local navigation process of each agent as an optimization problem and define an energy function that considers collision avoidance, attraction, velocity continuity, and direction control. Our energy function tries to match the results with real-world data characteristics. At a given moment, each agent chooses a velocity from a dataset. We align the control directions between simulation agents and real-world agents to diversify agents' possible behaviors and movements where there is relatively less input data available. To accelerate the computation, we utilize spatial continuity to reduce the number of possible collisions and use the velocity continuity to reduce the solution space for energy functions.

Overall, the novel contributions of our work include: 1. A general, optimization-based method to simulate heterogeneous multi-agent systems. We use our approach to simulate crowds, traffic, and any combination of those agents. In addition, we use a data-driven scheme to improve the plausibility of our simulation.
2. Two fast search methods. To achieve real-time simulations, we utilize spatial continuity and velocity continuity to search for possible collision-free solutions. In practice, these search methods result in a 32,298x speedup with 4,000 agents.
3. Ability to simulate new scenarios beyond the input datasets: Our method can simulate agent behaviors in a dense scenario, even if the original datasets correspond to sparse scenarios. Furthermore, our method can simulate behaviors of agents that may differ from those captured by the input data. We can also use direction control, which computes ideal directions, to guide agents in various environments.

We highlight the performance of our approach on different scenarios in Fig.~\ref{fig:resultsAll}. In practice, our approach can generate plausible trajectories and behaviors for tens or hundreds of heterogeneous agents at interactive rates. To demonstrate the benefits of our method, we have conducted two user studies to evaluate the benefits of our method over prior methods while using a top-down view and the agent's view. In both studies, participants exhibit significant preference for our method over a prior crowd simulation method~\cite{Karamouzas2017implicit} and a traffic simulation method~\cite{chao2018realistic}. We also conduct a user study to compare the user experience via VR and via desktop, and VR shows a better user experience (see Sec.~\ref{sec:userStudy}).

\section{Related Work}
There is considerable research in multi-agent simulation, including many algorithms for simulating crowds and traffic. In this section, we give a brief overview of prior methods for parameter estimation and data-driven simulation.

\subsection{Parameter Estimation and Real-World Characteristics}
Parameter estimation with real-world datasets improves the accuracy of simulation methods. Researchers utilize empirical data to compute the parameters used for rule-based or physically-based multi-agent simulation methods automatically. Wolinski et al.~\cite{wolinski2014parameter} present a method to compute optimal parameters for rule-based or physically-based multi-agent simulation algorithms. Berseth et al.~\cite{berseth2014steerfit} present an approach that computes parameters for steering methods by minimizing any combination of performance metrics. Karamouzas et al.~\cite{karamouzas2012simulating} use distortion and longitudinal dispersion of the group to evaluate the results from simulations. Our approach based on data-driven optimization is quite different from these methods.

Many techniques have been proposed to learn agent characteristics from empirical data and to then use them for multi-agent simulation. Lee et al.~\cite{lee2007group} present a crowd simulation method which use an agent model generated from real-world observations. Chao et al.~\cite{chao2013video-based} apply characteristics of drivers from an empirical video to an agent-based model. Boatright et al.~\cite{boatright2013context} classify the contexts and learn the characteristics from a dataset. Charalambous et al.~\cite{charalambous2014pag} present a real-time synthesis method for crowd steering behaviors with the temporal perception pattern. Bi et al.~\cite{bi2016data-driven} simulate the process of lane-changing in traffic by learning characteristics from features of real vehicle trajectories. Kim et al.~\cite{kim2016interactive} compute collision-free trajectories of virtual pedestrians by learning pedestrian dynamics from 2D trajectories. Our data-driven optimization algorithm is complimentary to these algorithms and can be combined with them.

Reconstruction of certain aspects of real-world scenes has also been used for multi-agent simulation, especially traffic simulation. Li et al.~\cite{li2017city-scale} reconstruct traffic with GPS mobile vehicle data. Wilkie et al.~\cite{wilkie2013flow} drive an agent-based traffic simulator by using the state of traffic flow estimated from sparse sensor measurements. Our approach is more general than these prior methods. Qiao et al.~\cite{qiao2018role} present a trajectory interpolation method by combining trajectory estimation and global optimization.

\subsection{Data-Driven Multi-Agent Simulation}
Patch-based methods transfer the original trajectories from empirical data into patches and connect these patches with some rules. Yersin et al.~\cite{yersin2009crowd} extend the concept of motion patches to dense populations in large environments. Li et al.~\cite{li2012cloning} animate large crowds with examples of multi-agent motions by using a copy-and-paste technique. Hyun et al.~\cite{hyun2013tiling} tile deformable motion patches, which describe episodes of the movements of multiple characters. Jordao et al.~\cite{jordao2014crowd} propose a crowd sculpting method to guide crowd motion by using intuitive deformation gestures.

As with patch-based methods, researchers replicate trajectory tubes extracted from empirical data to synthesize new agent animations. Lai et al.~\cite{lai2005group} introduce group motion graphs to animate groups of discrete agents with empirical data. Lerner et al.~\cite{lerner2007crowds} generate seemingly natural behaviors by copying trajectories from real people and applying them to simulated agents. Ju et al.~\cite{ju2010morphable} generate new animations, which can include arbitrary numbers of agents, by blending existing data. Zhao et al.~\cite{zhao2013data} cluster the examples extracted from human motion data and combine similar examples to produce an output. Li et al.~\cite{li2015biologically-inspired} propose a general, biologically-inspired framework with a three-level method using statistical information from real datasets. A new data-driven method has been proposed by Chao et al.~\cite{chao2018realistic}. They compute the velocity for each agent in each frame from empirical data. However, this method is time-consuming because it tries to minimize the overall traffic texture energy and is therefore not useful for interactive applications. Our approach is complimentary to prior data-driven  methods and presents a new method that combines data-driven with physics-based multi-agent methods.

\section{Data-Driven Optimization}
In this section, we introduce our data-driven optimization approach to simulating heterogeneous multi-agent simulations.

\subsection{Terminology and Notation}
We use \emph{agent} to represent the virtual character in our method. We also use the term \emph{state} to represent the motion characteristics of each agent. Our method is general and applicable for both 2D and 3D motions. State can therefore refer to an agent's movements in either 2D or 3D space. In this paper, we limit our discussions to 2D agents.

We use set $\mathcal{G}$ to specify the set of agents in the scenario. We use the vector $\mathbf{s}=[\mathbf{p},\mathbf{v},\mathbf{v}^{\rm d}]^{\rm T}$, $\mathbf{s} \in \mathbb{R}^6$ to specify an agent's state, where $\mathbf{p}$ is the agent's position, $\mathbf{v}$ is the velocity, and $\mathbf{v}^{\rm d}$ is the control direction that guides the motion direction of agents. Distinct from the velocity $\mathbf{v}$, the control direction $\mathbf{v}^{\rm d}$ controls the agent's global direction. We use $\hat{\mathbf{v}}=\frac{\mathbf{v}}{\|\mathbf{v}\|}$ to represent the unit vector of $\mathbf{v}$. We also use $\mathbf{v}_{i,n}$ to represent the velocity of agent $i$ at time $t_{n}$. The overall state of the group becomes $\mathcal{S}=\cup_{i}\mathbf{s}_{i}$. For any state $\mathbf{s} = [\mathbf{p},\mathbf{v},\mathbf{v}^d]\in \mathcal{S}$, $\mathbf{p}\in \mathcal{S}_{\rm p}$, $\mathbf{v}\in \mathcal{S}_{\rm v}$, $\mathbf{v}^{\rm d}\in \mathcal{S}_{\rm v^d}$. We represent our method by $\mathbf{M} = [S(), D(), I(), F()]^{\rm T}$, where $S$ is the environment evolution function, $D$ is the data processing function, $I$ is the initialization function, and $F$ is the decision making function. $S$ determines the external environment, which consists of the static environment (static obstacles, ground, etc.) and the dynamic environment (moving stimulus). $D$ processes the data set by transferring the trajectories to the estimated states $\mathcal{D}=\cup_n\mathcal{S}^*_n=\cup_n\cup_i\mathbf{s}_{i,n}^{*}$, where $\mathbf{s}_{i,n}^*=[\mathbf{p}_{i,n}^*,\mathbf{v}_{i,n}^*,\mathbf{v}_{i,n}^{d*}]$ denotes the state of agent $i$ at time $t_n$ of the dataset. For any $\mathbf{s}=[\mathbf{p},\mathbf{v},\mathbf{v}^d]\in \mathcal{D}$, $\mathbf{p}\in \mathcal{D}_{\rm p}$, $\mathbf{v}\in \mathcal{D}_{\rm v}$, $\mathbf{v}^d\in \mathcal{D}_{\rm v^d}$. $I$ initializes each agent's state: position, velocity, and control direction. $F$ is the main routine corresponding to our algorithm and computes a new state for each agent at each timestep.

\subsection{Overall Approach}
\begin{figure}[htb]
\centering
\includegraphics[width=\columnwidth]{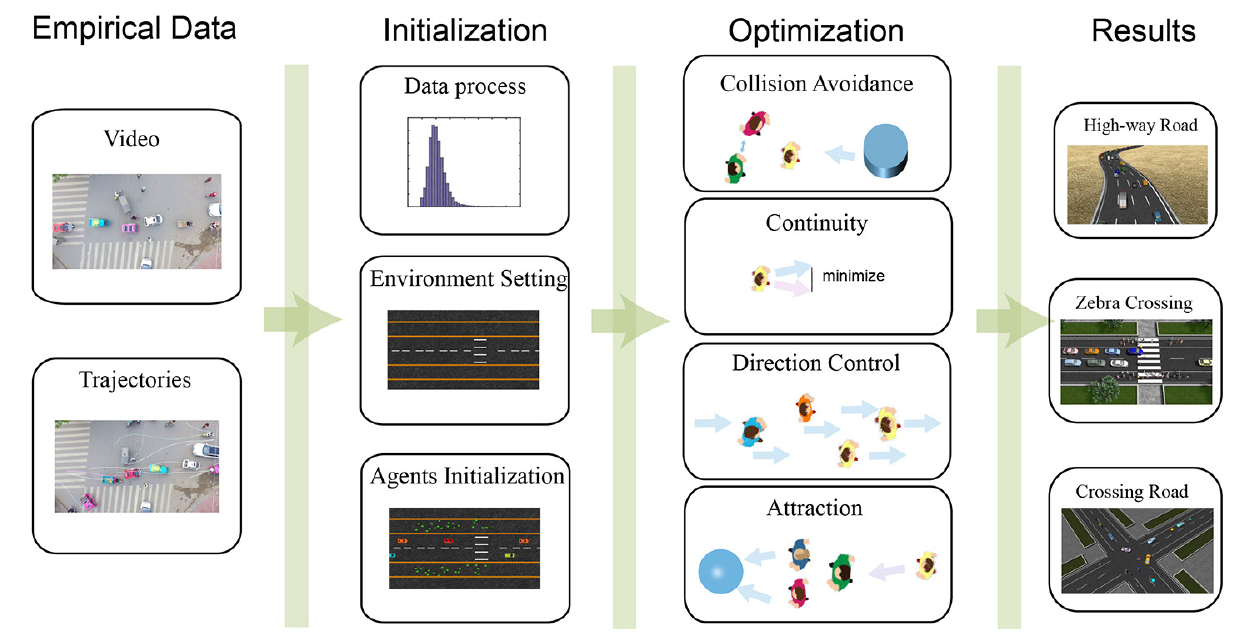}
\caption{
Overview of our data-driven model for simulating heterogeneous multi-agent systems. We highlight different components of our algorithm. The input empirical data can be videos from a top-down view or trajectories of agents. In the initialization, we first transfer real-world data into a consistent format. With the data and environment information set by the users, we initialize the positions and velocities for agents. We treat the motion decision-making or local navigation process of each agent at every timestep as an optimization problem, and the energy function takes into consideration several factors: continuity, collision avoidance, attraction, direction control, and any other constraints defined by users. Our model can simulate heterogeneous agents in the same scenario, including crowds, traffic, any combination of these agents, etc.
}
\label{fig:overview}
\end{figure}

Our model for simulating heterogeneous multi-agent systems references the datasets to control the trajectories and behaviors of the agents (see Fig.~\ref{fig:overview}). The datasets might be videos or other data representations, including trajectories or higher order features. We deal with different types of datasets and transform them into a unified representation, classifying the data by the magnitude of the velocity. The environment may also consist of static and dynamic obstacles. We initialize the position of each agent in the scene randomly and choose an initial velocity for each agent from our datasets. At each step of our simulator, we use an interactive optimization algorithm to make decisions for each agent. In particular, we solve this optimization problem by choosing a velocity from the datasets that tends to minimize our energy function. The energy function is defined based on the locomotion or dynamics rules of heterogeneous agents, including continuity of velocity, collision avoidance, attraction, direction control, and other constraints defined by users. In addition, our approach is general and can deal with different kinds of agents in the same way. We can capture corresponding motion characteristics with different datasets. As a result, we can simulate heterogeneous agents in the same physical space.

\subsection{Dynamics Computation}
An agent moves according to its surroundings, which include the other agents and the external environment (attractions, obstacles, roads, etc.). In these complex surroundings, the agent makes decisions in relation to all these elements. At each timestep, the state of each agent can be computed as
\begin{equation}\label{formula:stateTransfer}
\mathbf{s}_{i,n+1} = F(t_{n},i,\mathcal{S}_{t_{n}}, S(t,\mathbf{p}_{i,0}),\mathcal{D}); \mathbf{s}_{i,0} = I(i,S(t_{0},\mathbf{p}_{i,0}),\mathcal{D}).
\end{equation}
Because the external environments may be time-varying, we set the environment evolution function as a function of time. In addition, our method is data-driven, so the function $F()$ is also a function of $\mathcal{D}$.

We expand Eq.~\ref{formula:stateTransfer} to a system of equations, and obtain
\begin{equation}\label{formula:updateEq}
\begin{matrix*}[r]
\mathbf{p}_{i,n+1} = \mathbf{p}_{i,n} + \mathbf{v}_{i,n+1}\Delta t,\\
\\\mathbf{v}_{i,n+1} = \underset{\mathbf{v}\in\mathcal{D}_{\rm v}}{\operatorname{argmin}}\ E(t_{n},i,\mathbf{v},\mathcal{S}_{n}, S(t_{n},\mathbf{p}_{i,n}),\mathbf{v}^{\rm d}_{i,n+1}),
\\\mathbf{v}^{\rm d}_{i,n+1} = R(\mathbf{p}_{i,n},S(t_{n},\mathbf{p}_{i,n})),
\end{matrix*}
\end{equation}
where $E(t_{n},i,\mathbf{v},\mathcal{S}_{t}, S(t,\mathbf{p}_{i,t}),\mathbf{v}^{\rm d}_{i,n+1})$ is the energy function that chooses the optimal velocity for agent $i$ at time $t_{n+1}$. $R(\mathbf{p}_{i,n},S(t_{n},\mathbf{p}_{i,n}))$ is a function that computes the control direction $\mathbf{v}^{\rm d}$ for each agent at each time. We compute a velocity that minimize the energy function. If we search the velocity from a continuous-space, our method becomes an energy-based model. In order to capture the characteristics of different kinds of agents easily, we search for the velocity from the states in the dataset $\mathcal{D}$, which belongs to a discrete space. If the states generated from the dataset are unlimited, the simulation results will approximate the simulation results generated from the method with the continuous velocity space .

To simulate heterogeneous agents in the same physical space, we consider the common locomotion rules of multi-agent systems for the energy function $E(t_{n},i,\mathbf{v},\mathcal{S}_{n}, S(t_{n},\mathbf{p}_{i,n}),\mathbf{v}^{\rm d}_{i,n+1})$ including collision avoidance, attraction, continuity, direction control, and any other constraints.
\begin{equation}\label{formula:Energy}
E(t_{n},i,\mathbf{v},\mathcal{S}_{n}, S(t_{n},\mathbf{p}_{i,n}),\mathbf{v}^{\rm d}_{i,n+1}) = \sum_{k\in\mathcal{\theta}}{w_{k}E_{k}(t_{n},i,\mathbf{v},\mathcal{S}_{n}, S(t,\mathbf{p}_{i,n}),\mathbf{v}^{\rm d}_{i,n+1})},
\end{equation}
where $\theta = \{{\rm m,c,a,d,s}\}$, $E_{\rm m}$ is the energy for velocity continuity, $E_{\rm c}$ is the energy for collision avoidance, $E_{\rm a}$ is the energy for attraction, $E_{\rm d}$ is the energy for direction control, and $E_{\rm s}$ is the energy function for constraints of certain kinds of agents. $w_{\rm m}$, $w_{\rm a}$, $w_{\rm t}$, $w_{\rm d}$, and $w_{\rm s}$ are the weights of these terms, respectively. Velocity continuity is used to ensure that the agents move smoothly. Collision avoidance is a crucial part of multi-agent simulation. Attraction helps agents remain cohesive with other agents in the same group and has been widely used in multi-agent simulation literature~\cite{Boids87Flocks}. The direction control represents the direction preference for agents according to the environment. These four elements can describe the basic factors considered by agents when moving. It is possible to add more constraints to control the movements of agents in $E_{\rm s}$.

\subsection{Continuity}
Because of the physical limitations, agents cannot change their motion states frequently or abruptly within a $\Delta t$ time. Thus, the agent $i$ has a tendency to choose a velocity close to $\mathbf{v}_{i,t}$ at a time $t+1$. The continuity energy is used to indicate that the agent tends to keep its velocity unchanged to save its overall energy:
\begin{equation}\label{formula:Continuity}
E_{\rm m} = w_{m1}E_{\rm m}^{\rm {dir}} + w_{m2}E_{\rm m}^{\rm L},
\end{equation}
where $E_{\rm m}^{\rm {dir}} = \left \| \hat{\mathbf{v}}_{i,v}- \hat{\mathbf{v}} \right \|_2$ is for direction continuity and $E_{\rm m}^{\rm L}=\left \| \| \mathbf{v}_{i,n}\|- \| \mathbf{v} \| \right \|_2$ is for velocity length continuity. $\mathbf{v}_{i,n}$ is the velocity of agent $i$ at time $t_{n}$.

\subsection{Collision Avoidance}
Collision avoidance is a major issue in multi-agent simulation~\cite{Karamouzas2014Universal,wolinski2016warpdriver}. To avoid collisions with other agents or the environmental obstacles in the scene, the agent should choose a velocity that will not cause a collision after one of more timesteps by assuming that all objects keep moving with their current velocities. Here, we consider two kinds of collisions to avoid: instantaneous collisions and anticipatory collisions.
\begin{equation}\label{formula:Collision}
E_{\rm c} = w_{c1}E_{\rm c}^{\rm Ins} + w_{c2}E_{\rm c}^{\rm Anti},
\end{equation}
where instantaneous collision avoidance energy $E_{\rm c}^{\rm Ins}$ only considers the possible collisions after a timestep, and anticipatory collision energy $E_{\rm c}^{\rm Anti}$ considers the possible collisions after anticipation time $T$.

Instantaneous collision avoidance energy $E_{\rm c}^{\rm Ins}$ is given as
\begin{equation}\label{formula:CollisionIns}
E_{\rm c}^{\rm Ins} = \frac{1}{|\Omega_{\rm c}(\Delta t,i,\mathbf{v},\mathcal{S}_{n}, S(t,\mathbf{p}_{i,n}))|}\sum_{Q \in \Omega_{\rm c}(\Delta t,i,\mathbf{v},\mathcal{S}_{n}, S(t_{n},\mathbf{p}_{i,n}))} e^{d_{\rm c}-d(\Delta t,\mathbf{s}_i,\mathbf{s}_Q,\mathbf{v})},
\end{equation}
where $\Omega_{\rm c}(\Delta t,i,\mathbf{v},\mathcal{S}_{n}, S(t_{n},\mathbf{p}_{i,n}))$ is the predicted neighborhood of agent $i$ after time $\delta t$. The neighborhood consists of agents that probably collide with agent $i$. $d(\Delta t,\mathbf{s}_i,\mathbf{s}_Q,\mathbf{v})$ is the predicted distance between agent $i$ and agent $Q$. For each agent, we only consider collision avoidance within a distance $d_{\rm c}$. Similarly, the anticipatory collision avoidance energy $E_{\rm c}^{\rm Anti}$ can be given as
\begin{equation}\label{formula:CollisionAnti}
\begin{split}
E_{\rm c}^{\rm Anti} = \frac{1}{|\Omega_{\rm c}(T\Delta t,i,\mathbf{v},\mathcal{S}_{n}, S(t,\mathbf{p}_{i,n}))|} \cdot\\
\sum_{Q \in \Omega_{\rm c}(T\Delta t,i,\mathbf{v},\mathcal{S}_{n}, S(t_{n},\mathbf{p}_{i,n}))} e^{d_{\rm c}-d(T\Delta t,\mathbf{s}_i,\mathbf{s}_Q,\mathbf{v})},
\end{split}
\end{equation}
where $\Omega_{\rm c}(T\Delta t,i,\mathbf{v},\mathcal{S}_{n}, S(t_{n},\mathbf{p}_{i,n}))$ is the predicted collision neighborhood of agent $i$ after time $T\Delta t$. $d(T\Delta t,\mathbf{s}_i,\mathbf{s}_Q,\mathbf{v})$ is the predicted distance between agent $i$ and agent $Q$ after time $T$.

\subsection{Attraction}
If the agents want to move together as a group, we need to account for some attraction forces between them. The agent therefore prefers to choose a velocity that brings it closer to a group, allowing it to become a part of the group over the next few frames. In addition, agents may also be attracted by external stimuli.
The attractions in our model include the attraction between the agents and the environment. The attraction energy is given as
\begin{equation}\label{formula:Attraction}
E_{\rm a} = \frac{1}{|\Omega_{\rm a}(\Delta t,i,\mathbf{v},\mathcal{S}_{n}, S(t_{n},\mathbf{p}_{i,n}))|}\sum_{Q \in \Omega_{\rm a}(\Delta t,i,\mathbf{v},\mathcal{S}_{n}, S(t_{n},\mathbf{p}_{i,n}))} d^2(\Delta t, \mathbf{s}_i, \mathbf{s}_Q,\mathbf{v}),
\end{equation}
where $\Omega_{\rm a}(\Delta t,i,\mathbf{v},\mathcal{S}_{n}, S(t_{n},\mathbf{p}_{i,n}))$ is the predicted attraction neighborhood of agent $i$ after time $\Delta t$.

\subsection{Direction Control}\label{sec:directionControl}
We use direction control to imitate agents moving toward their goals. In this case, the agents try to choose velocities that point to their goals or that parallel the path to their goals.
We assume that every agent has a goal position to guide its local movement. The goal might change over time. This goal can also be treated as a direction control defined by the users. The energy for direction control is presented as
\begin{equation}\label{formula:Control}
E_{\rm d} = \left \| \mathbf{v}^{\rm d} - \hat{\mathbf{v}}\right \|_2,
\end{equation}
where $\mathbf{v}^{\rm d}$ is the control direction.

\section{Multi-agent System Simulation with Data-Driven Optimization}
In this section, we present more details about our method, as it is used to simulate heterogeneous agents.
\subsection{State Estimation for the Dataset}
The dataset of our method consists of trajectories that are time series of positions, $\mathcal{L}: \mathbf{Y}_1,\mathbf{Y}_2,...,\mathbf{Y}_n... $. We estimate the state $\mathbf{s}^*_n=[\mathbf{p}^*_n,\mathbf{v}^*_n,\mathbf{v}^{\rm d*}_n]$ in the dataset based on these trajectories, and obtain the estimated position $\mathbf{p}^*_n = \mathbf{Y}_n$ and velocity $\mathbf{v}^*_n=\frac{\mathbf{Y}_n-\mathbf{Y}_{n-1}}{\Delta t}$. Estimating the control direction $\mathbf{v}^{\rm d*}_n$ is equivalent to estimating the direction to the corresponding agent's goal, according to Sec.~\ref{sec:directionControl}. Therefore, if the agent only moves one way in the scenario, it is in the control direction; if the agent changes its direction or goal in the dataset, we estimate its control direction at time $t$ by computing the direction of its displacement, $\mathbf{v}^{\rm d*} = \frac{\mathbf{Y}_n-\mathbf{Y}_{n-\delta}}{\| \mathbf{Y}_t-\mathbf{Y}_{t-\delta}\|}$, which is estimated every $\delta \Delta t$ time.

\subsection{Direction Adaptation to Different Scenarios}
According to Eq.~\ref{formula:updateEq}, if we directly search the optimal velocity for each agent from the dataset, the synthesized scenario will be limited in its ability to achieve plausible movements by the scenario of the dataset. To eliminate these constraints, we map the local coordinate of the dataset to that of the scenario in the simulation by align their forward directions. As a result, we can simulate scenarios that may be different from the dataset. We suppose that the simulated scenario and the dataset have the same relative position relationship between the direction of velocity and the control direction; that is, $\hat{\mathbf{v}}-\mathbf{v}^{\rm d} = \hat{\mathbf{v}}^*-\mathbf{v}^{\rm d*}$. Therefore, we obtain
$\mathbf{v} = \|\mathbf{v}^*\|(\mathbf{v}^{\rm d} + (\hat{\mathbf{v}}^*-\mathbf{v}^{\rm d*}))$.

\subsection{Distance and Neighborhood}
We hypothesize that the velocity of an agent remains unchanged over a short time. If the agent $i$ moves with the velocity $v$ chosen from the dataset, the predicted distance between agent $i$ and agent $Q$ after a short time $t$ becomes
\begin{equation}\label{formula:distanceOrigin}
d(t,\mathbf{s}_i,\mathbf{s}_Q,\mathbf{v}) = \mathbf{p}_{i} + \mathbf{v} t - (\mathbf{p}_{Q}+\mathbf{v}_{Q} t).
\end{equation}
In this equation, we assumed that the shapes of the two agents can be ignored. In fact, we cannot ignore the shapes of most of the agents or obstacles. Thus, we modify Eq.\ref{formula:distanceOrigin} and obtain
\begin{equation}\label{formula:distanceDef}
d(t,\mathbf{s}_i,\mathbf{s}_Q,\mathbf{v}) = \mathbf{p}_{i} + \mathbf{v} t - (\mathbf{p}_{Q}+\mathbf{v}_{Q} t) - (R_i^{\rm dir}+R_Q^{\rm dir}),
\end{equation}
where $R_i^{\rm dir}$ is the radius of agent $i$ in the direction toward agent $Q$. $R_Q^{\rm dir}$ is also a directional radius of agent $Q$. The shapes of different agents can be different. For example, we use a rectangle to represent a car and a circle to represent a pedestrian. If $Q$ is an identity in the environment, Eq.~\ref{formula:distanceDef} becomes a distance function between an agent and the identity in the environment.

In contrast to the existing methods~\cite{ren2017group}, the agents in our method try to avoid collisions with not only the homogeneous agents but also the heterogeneous agents. To avoid collisions, each agent tries to keep away from other agents or obstacles when they get too close. We define the neighborhood for collision avoidance as
\begin{equation}\label{formula:NeighborhoodC}
\Omega_{\rm c}(t,i,\mathbf{v},\mathcal{S}_{t}, S(t,\mathbf{p}_{i,t})) = \left\{Q\big|d(t,\mathbf{s}_i,\mathbf{s}_Q,\mathbf{v})<d_{\rm c}, Q \in \mathcal{G}\setminus\{i\} \cup \mathcal{G}_{\rm c} \right\},
\end{equation}
where $d_{\rm c}$ is the threshold distance for collision avoidance and $\mathcal{G}_{\rm c}$ is the set of obstacles in the scenario. Each agent considers collision avoidance with the agents or obstacles within a distance $d_{\rm c}$. Each agent tries to keep close to the agents in its group or to the external attraction stimulus if the distance between the agents is large. We define the neighborhood for attraction as
\begin{equation}\label{formula:NeighborhoodA}
\Omega_{\rm a}(t,i,\mathbf{v},\mathcal{S}_{t}, S(t,\mathbf{p}_{i,t})) = \left\{Q\big|d(t,\mathbf{s}_i,\mathbf{s}_Q,\mathbf{v})>d_{\rm a}, Q \in \mathcal{G} \cup \mathcal{G}_{\rm a} \right\},
\end{equation}
where $d_{\rm a}$ is the threshold distance for attraction and $\mathcal{G}_{\rm a}$ is the set of attraction in the scenario. An entity that is treated as an attraction can also be an obstacle if the shape of it cannot be ignored, that is, $\mathcal{G}_{\rm c} \cap \mathcal{G}_{\rm a} \neq \varnothing$.

\subsection{Faster Computation}
If we use a brute force method to solve Eq.~\ref{formula:updateEq}, the computation cost will be large. The underlying time complexity will be $O(n^2m)$ with $n$ agents in the simulation and $m$ estimate states in the dataset. The most time-consuming parts are searching for the optimal velocity from the dataset and finding the neighborhood for each agent. To achieve interactive performance, we propose two acceleration methods.

\subsubsection{Reduced Solution Space}
To find the optimal velocity for each agent efficiently, we reduce the solution space of Eq.~\ref{formula:updateEq}. We classify the estimated states of the dataset into groups based on the magnitude of the velocity. Considering the continuity of motion, we search for the velocity for each agent in the current group of velocities and in the adjacent group,
\begin{equation}\label{formula:VelSpaceObs}
\mathbf{v}_{i,n+1}\in \bigcup_{m=l-z}^{l+z}\{\mathbf{v}^{n*}\},
\end{equation}
where $\{\mathbf{v}^{l}\}$ is the set of velocities of the group $l$ to which $\mathbf{v}_{i,t}$ belongs, $z$ is the scope of the number of groups that are considered for computing optimal velocity, and the group $\{\mathbf{v}^{m}\}$ with $m\in[l-z,l+z]$ is the neighborhood of $\{\mathbf{v}^{l}\}$.

\subsubsection{Grid in Space}
To reduce the time consumption for computing the neighborhood for each agent, we introduce the idea of a grid in space from fluid simulation~\cite{bridson2007fluid}. For our simulation, the 2D plane is divided into 2D grids. We suppose that $\mathcal{O}_{x,y}$ denotes the set of all agents in the grid $O_{x,y}$. Then the candidate neighborhood of $i$ in grid $O_{x,y}$ is reduced from $\mathcal{G}$ to $\mathcal{G}'$,
\begin{equation}
\mathcal{G}' = \bigcup_{k_1=x-1}^{x+1}\bigcup_{k_2=y-1}^{y+1} \mathcal{O}_{k_1,k_2}.
\end{equation}
When we search the neighborhood for collision avoidance or attraction, we compare the distances of the agents in the grid $O_{x,y}$ with the agents adjacent to this grid instead of comparing them to all the agents in the scenario.

\section{Results}
In this section, we highlight the performance of our approach in generating simulations of crowds, traffic, and combinations of different types of agents. We have implemented our approach in C++ on a desktop machine with a 3.30GHz Inter Xeon CPU E3-1230 v3 4-core processor and 32GB memory. The performances for different scenarios are given in Table~\ref{tab:runTime}. Table~\ref{tab:weights} shows the weights of all the benchmarks. We define the user control for each pedestrian with speed control $E_{\rm a} = E_{\rm SG} = \| \|\mathbf{v}\| - v_i\|$, where $v_i$ is the ideal speed for agent $i$. We define the user control for each car with lane control $E_{\rm a} = E_{\rm Cons} = |\mathbf{v} \cdot (\mathbf{v}^{\rm d})^\perp|$. Cars try to drive in the middle of the lane.

\subsection{Data Acquisition}
Our method accepts different kinds of input datasets if those datasets contain the velocity information for the agents. Any form of discontinuity or a small amount of abnormal data in the datasets is acceptable.

In our current framework, we have used some widely available datasets from different scenarios and environments. The datasets for crowd simulation include two scenarios: one is from~\cite{zhang2012ordering} and features two-dimensional bidirectional movements with 304 pedestrians and 1,273 frames; the second is from~\cite{lerner2007crowds} and features street scenarios with 8-148 pedestrians and 9,014 frames. We set the control directions for the first dataset as the directions that point to the agents' destinations. For the second dataset, the control direction of one agent at a certain frame is the direction that points from its current position to its position after 10 frames.

The traffic dataset is extracted from the Next Generation Simulation (NGSIM) datasets~\cite{NGSIM}, which include detailed, high-quality highway traffic datasets. We extract 300 frames and 161 cars in total. We set the direction of the road as the estimation of the control directions of the cars. The datasets corresponding to the mixed traffic scenarios (including pedestrians, bicycles, tricycles, and cars) are generated from videos. The video was recorded in Shandong, China. We use the optical flow tracking method~\cite{horn1981determining} to trace the agents. The extracted data consists of 435 frames and contains 3 pedestrians, 10 bicycles, 10 tricycles, and 2 cars. The control direction for each agent in every frame is computed by averaging the directions of the agent from 30 frames.

\subsection{Human Crowd}
We simulate three benchmark scenarios with crowds representing each pedestrian as a circle.

\begin{figure}[htb]
\includegraphics[width=\columnwidth]{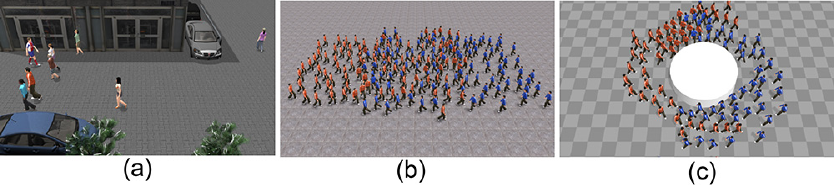}
\caption{
The mixed crowds with different control directions. (a) Pedestrians with changing control directions walk on a street. (b) Two crowds with inverse control directions. The pedestrians with the same clothes represent individuals in the same crowd. The crowds walk to their own destinations while avoiding collisions with each other. (c) We add an obstacle to the scenario. In addition to avoiding collisions with each other, crowds should also avoid collisions with this obstacle.}
\label{fig:crowd}
\end{figure}

\textbf{Crowd-1:} We simulate behaviors of pedestrians on a street with the dataset from~\cite{lerner2007crowds} to show that our method can reproduce a scenario from the dataset. In this scenario, we set the number of agents in the initialization and control directions to be the same as in the dataset. Pedestrian agents, represented as circles, mainly avoid collisions with other pedestrians that are close to them in the scene (see Fig.~\ref{fig:crowd}(a)).

\textbf{Crowd-2:} In this scenario, we simulate two groups (50 pedestrians in each group) with control directions inverse to those from the dataset~\cite{zhang2012ordering}. We randomly locate the agents in each group at one side of the road and randomly choose a velocity for each agent from the dataset in the initialization. The control direction points from the agent's position to the agent's goal on the other side of the road. The reference speed is the magnitude of the initial velocity. Agents are attracted to those in the same group and avoid collisions with other agents, including pedestrians in other groups (see Fig.~\ref{fig:crowd}(b)).

\textbf{Crowd-3:} Based on the benchmark \textbf{Crowd-1}, we add a cylindrical obstacle in the center of the road (see Fig.~\ref{fig:crowd}(c)). We also use the dataset~\cite{zhang2012ordering} in this benchmark. The initialization method for this benchmark is the same as for the benchmark Crowd-2. In our simulations, we set different control directions for different groups and agents in the same group share the same control direction. Agents avoid the obstacle like they avoid other agents.

\subsection{Traffic}
In traffic simulations, vehicle-agents mainly interact with the cars that are adjacent to them in the same lane, avoiding collisions when they are too close and being attracted by the leader cars when the distance to that car becomes too large. However, cars that are changing lanes also interact with the adjacent cars in the target lanes. The control directions for the cars in traffic are the directions of the lanes to which they currently belong.

\begin{figure}[htb]
\includegraphics[width=\columnwidth]{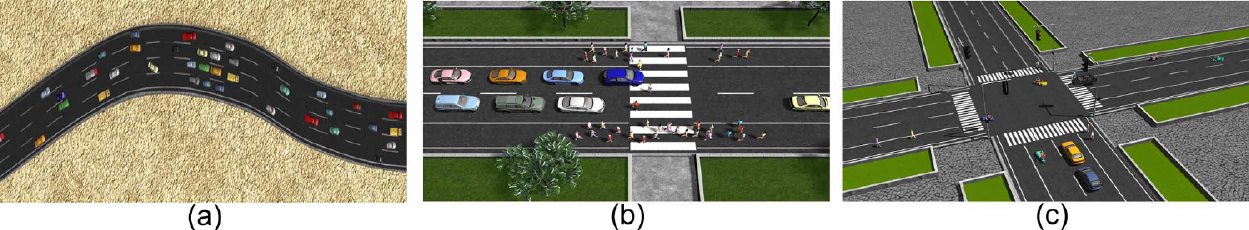}
\caption{Traffic simulation. (a) Traffic on a twisting 4-lane highway. (b) A combination of cars and crowds. Some pedestrians are walking on the sidewalk. Cars can be treated as obstacles for crowds and vice versa. (c) Congested traffic in an urban crossroad with a traffic lights.
}
\label{fig:traffic}
\end{figure}

\textbf{Traffic-1:} With our method, we can simulate traffic on twisting roads with the straight high way traffic dataset~\cite{NGSIM} (see Fig.~\ref{fig:traffic}(a)). During the initialization step, 80 cars are distributed on the road. The distance between two adjacent cars is chosen randomly from the dataset. We also randomly select the magnitude of the velocity for each agent from the dataset, and the direction of the velocity is the same as the direction of the road on which the agent is driving. The control direction of each agent is always the direction of the road. In this benchmark, the directions of agents in different positions on the twisting road vary.

Our method is general, so we can mix different kinds of agents in the same scenario. In this section, we show two benchmarks: a zebra striped crosswalk and a crossroad without traffic lights.

\textbf{Traffic-2:} In this benchmark, we simulate a case in which people want to cross the road (see Fig.~\ref{fig:traffic}(b)). We use dataset~\cite{zhang2012ordering} for the crowd and dataset~\cite{NGSIM} for the traffic. Each pedestrian has a certain possibility of crossing the road. Once the pedestrian starts to cross road, the control direction becomes perpendicular to the road direction and the pedestrian needs to avoid not only other pedestrians, but also the cars around it. At the same time, the surrounding cars need to stop if the pedestrian is in front of them, and the attractive force from the leading cars disappears for these cars. We implement these interactions by adding corresponding objects to the interaction domain of agents.

\textbf{Traffic-3:} Our model can handle congested scenarios with different or heterogeneous agents. Here we simulate agents (8 pedestrians, 8 bicycles, 8 tricycles, and 8 cars) crossing a congested road with a traffic light (see Fig.~\ref{fig:traffic}(c)). We classify the dataset into groups according to the corresponding type of agent in the original data and choose the velocities of the agents from the corresponding class. Furthermore, we classify the four kinds of agents into two types with different motion constraints. The first type includes pedestrians and bicycles, which can overtake the agents in front of them in the same lane. The second type includes tricycles and cars, which cannot overtake the agent in front in the same lane. When an agent reaches the crossing, the control direction is the interpolation of the original road direction and the target road direction. The rule for traffic light is not strictly same as that in the real world.

\begin{table*}[htb]
\centering
\begin{tabular}{|c|c|c|c|c|c|c|}
  \hline
  Scenario & Types & Behavior &$N$ & Dataset & Time(s/f) \\
  \hline\hline
   Crowd-1& human& walking on street& 8-148 & [Lerner et al. 2007]& 0-0.0040  \\
   \hline
   Crowd-2& human& mixture of two crowds& 100 &[Zhang et al. 2012]& 0.0209 \\
   \hline
   Crowd-3& human& avoiding static obstacles & 79 &[Zhang et al. 2012]& 0.0192  \\
   \hline
   Traffic-1& car& movements on a twist road& 80 & [NGS 2013]   & 0.0137 \\
   \hline
   Traffic-2& human/car& movements on a crossing road& 30/35 &[NGS 2013]/[Zhang et al. 2012]& 0.0378 \\
   \hline
   Traffic-3&\makecell{human/bicycle\\/tricycle/car}&mixture of multiple systems & \makecell{8/8\\/8/8} & video from Shandong, China & 0.0028 \\
   \hline
\end{tabular}
\caption{Performance for different scenarios. We summarize the characteristics of the simulation scenarios in this paper. The agents include humans, cars, bicycles, and tricycles. The datasets used for input data vary. We use seconds per frame to measure the time performance of the simulations. Our method can achieve realtime performance using 4 cores on a CPU.}
\label{tab:runTime}
\end{table*}

\begin{table*}[htb]
\centering
\begin{tabular}{|c|c|c|c|c|c|c|c|c|c|}
\hline
\multicolumn{2}{|c|}{Scenario}          & $E_{\rm t}^{\rm {dir}}$ & $E_{\rm t}^{\rm L}$ & $E_{\rm c}^{\rm Ins}$ & $E_{\rm c}^{\rm Anti}$ & $E_{\rm a}$ & $E_{\rm d}$ & $E_{\rm Cons}$ & $E_{\rm SG}$ \\ \hline
\multicolumn{2}{|c|}{Crowd-1}           & 1.0                     & 1.0                   & 1.0                     & 1.0                      & 0           &1.0          & 0              & 0.5          \\ \hline
\multicolumn{2}{|c|}{Crowd-2}           & 1.0                     & 1.0                   & 1.0                     & 1.0                      & 0           &1.0          & 0              & 1.5          \\ \hline
\multicolumn{2}{|c|}{Crowd-3}           & 0.83                     & 1.0                   & 0.67                     & 0.67                      & 0.0            &0.83          & 0.0              & 1.0           \\ \hline
\multicolumn{2}{|c|}{Traffic-1}         & 0.5                       & 0.5                   & 1.0                     & 1.0                      & 2.0           & 3.0           & 10.0             &10.0            \\ \hline
\multirow{2}{*}{Traffic-2} & Pedestrian & 1.0                       & 1.0                   & 1.0                   & 1.0                    & 0           & 1.5           & 1.0              & 10.0            \\ \cline{2-10}
                           & Car        & 5.0                       & 1.0                   & 1.0                     & 1.0                      & 2.0           & 5.0           & 1.0              & 10.0            \\ \hline
\multirow{2}{*}{Traffic-3} & Type-1     & 10.0                      & 1.0                   & 1.0                     & 1.0                      & 0           & 5.0           & 10.0             & 5.0            \\ \cline{2-10}
                           & Type-2     & 0.5                       & 0.5                   & 1.0                     & 1.0                      & 2.0           & 3.0           & 1.0              & 10.0           \\ \hline
\end{tabular}
\caption{The weights. This table gives the weights for the direction continuity $E_{\rm t}^{\rm {dir}}$, the speed continuity $E_{\rm t}^{\rm L}$, instantaneous collision avoidance $E_{\rm c}^{\rm Ins}$, anticipated collision avoidance $E_{\rm c}^{\rm Anti}$, attraction $E_{\rm a}$, direction control $E_{\rm d}$, position constraint $E_{\rm Cons}$, and speed control $E_{\rm SG}$ in each scenario.}
\label{tab:weights}
\end{table*}

\subsection{VR scenarios}
\begin{figure}[htb]
\centering
\includegraphics[width=\columnwidth]{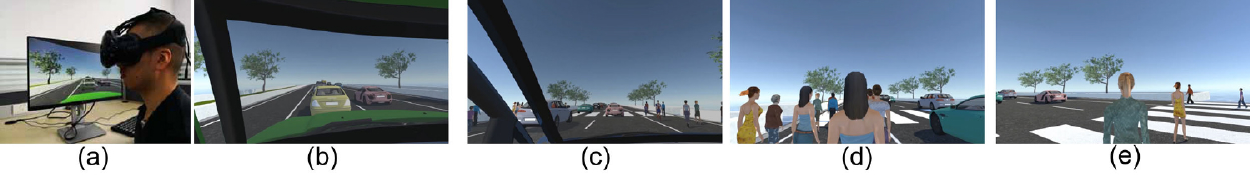}
\caption{
The avatar in a VR scenario. (a) We provide the user with an immersive VR experience from a first-person perspective with HTC Vive. (b) The avatar drives a car on an high-way road. (c) The avatar drives a car on an urban traffic road. (d) The avatar is walking on the sideroad. (e) The avatar is walking on the crosswalk.
}
\label{fig:VR}
\end{figure}

Our method can be applied to VR scenarios. We model the user as an avatar in the VR scenario with a first-person perspective (see Fig.~\ref{fig:VR}) (a). The user can sit in a car and observe the movements of other cars around it (see Fig.~\ref{fig:VR} (b) and (c)). As a walker, the user can also see the traffic flow and other pedestrians at the roadside (see Fig.~\ref{fig:VR} (d) and (e)).

\section{Time Performance}


To test the time performance of our method, we simulate a crowd in a scenario with the size of 1,000*1,000. There is no obstacle in the scenario. During the initialization, we randomly locate $N$ agents at random positions. The initial velocities of the agents are randomly copied from the dataset~\cite{zhang2012ordering}. We set the grid size of the simulation as 10, and the $z$ for Eq.~\ref{formula:VelSpaceObs} as 2.

\begin{figure}[htb]
\centering
\includegraphics[width=\columnwidth]{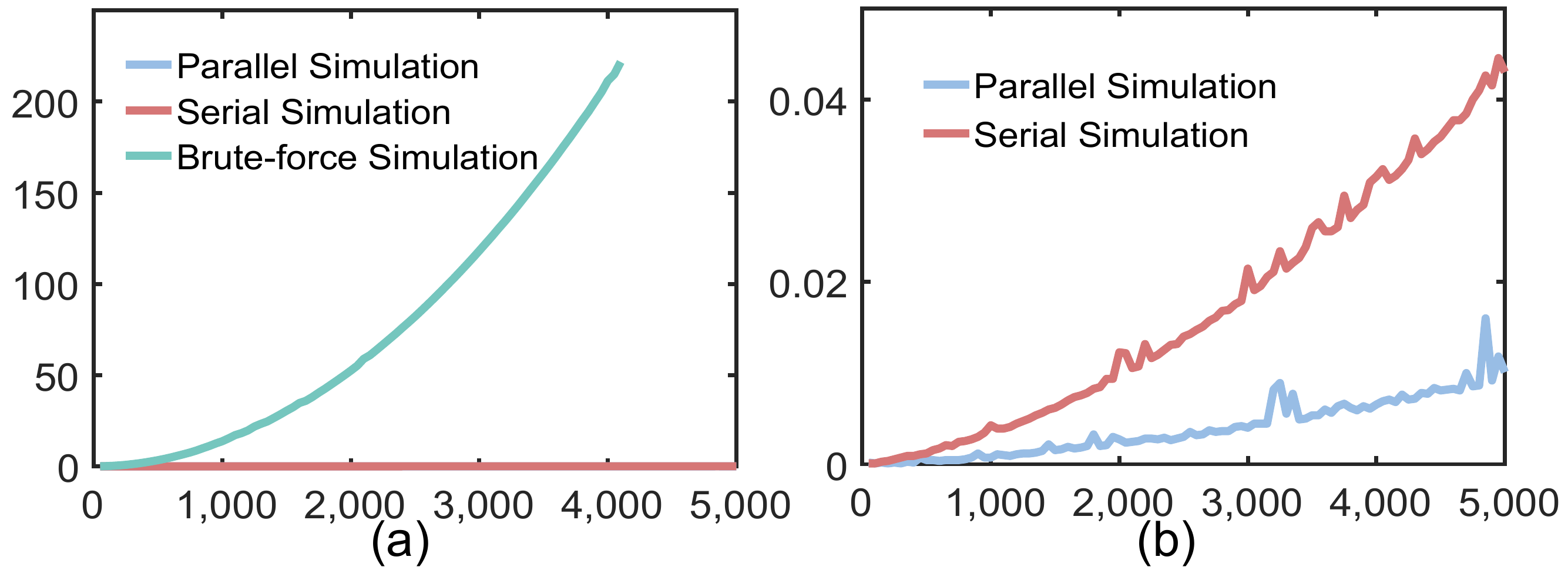}
\caption{
Time performance. We take a crowd as an example to analyze the time performance of the simulation. (a) We compare the time performance of the brute-force method and our method. With our two search methods, we can improve the performance with 32,298x speedup for 4000 agents. (b) We compare the performance of an 8-threaded parallel implementation with a single-threaded implementation. With parallel computing, as the number of agents increases, the simulation time increases much more slowly. Our method can even simulate 5,000 agents in realtime on a PC machine with a 4.00GHz Intel i7-6700k CPU processor and 16GB memory.}
\label{fig:timePerformance}
\end{figure}

In our method, we utilize spatial continuity and velocity continuity to reduce possible collisions among the agents. We use the size of the solution space of the optimization function in Eq.~\ref{formula:updateEq} to improve the runtime performance of our simulation. We divide the space into grids and each grid records the agents that belong to it. When we search for the neighbors of each agent, we only need to search the grid to which the agent belongs and the grids that are adjacent to this grid. As a result, our method can reduce the time consumption for multi-agent simulations dramatically (see Fig.~\ref{fig:timePerformance} (a)).

Because we can solve the optimization problem for each agent at the same time, we can also easily parallelize our method. Taking the crowd as an example, we compare the time complexity of our simulation using a serial implementation against a parallel implementation (see Fig.~\ref{fig:timePerformance} (b)). Our parallel implementation can simulate more than 5,000 agents in realtime on a multi-core processor with four cores.

To evaluate the performance of our method further, we compute the running time (seconds per frame) of all the simulation results mentioned in this paper (see Table~\ref{tab:runTime}). Our method can achieve real-time performance in various scenarios with multiple kinds of input dataset. The time complexity is not only related to the number of agents in the simulation, but also to the number of classes and the number of data points in each dataset. As a result, similar scenarios with the same number of agents may have different time performances.

\section{User Studies and Evaluation}\label{sec:userStudy}
We conduct two user studies to evaluate the plausibility of our method and one user study to show a better user experience through VR. The eight cases in the first user study are conducted from an overhead view to show the agents' movements. In the second user study, we adopt the agent's view in each case, meaning that the view is closer to that of a participant in his/her daily life. In the third user study, we compare the results as shown in immersive VR and those shown on a desktop in four different scenarios or agents' views.

\begin{figure*}[htb]
\centering
\includegraphics[width=2\columnwidth]{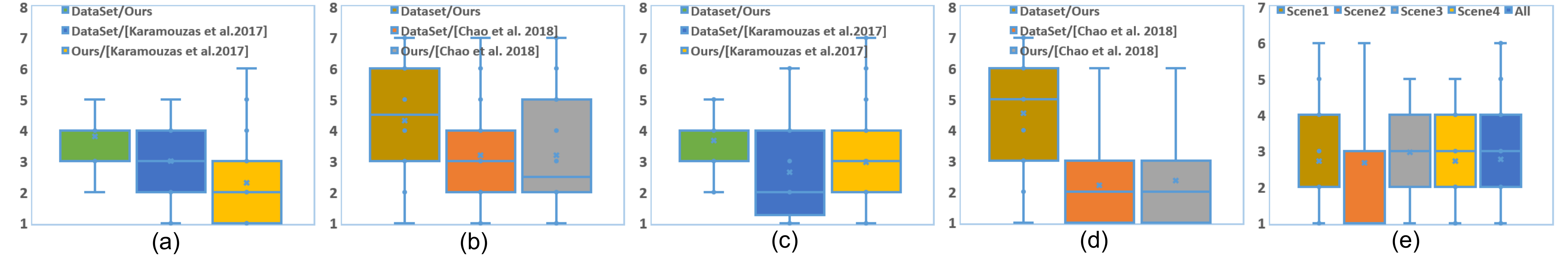}
\caption{
Plausibility scores of the user study. We use a 7-point Likert scale to measure the plausibility of the methods. The lower the score, the more the participants prefer the method on the left; the higher the score, the more the participants prefer the method on the right. (a)The statistics for crowd simulation with an overhead view. Participants cannot tell the difference between the dataset and our method. Compared to method~\cite{Karamouzas2017implicit}, the participants think the results of our method are more plausible. (b) The statistics for traffic simulation with an overhead view. Our method gets a higher score than method~\cite{Karamouzas2017implicit} when compared with the dataset. We also get better results in the user study with the dataset. (c) The statistics for crowd simulation from an agent view. Our method is closer to the dataset. The participants believe that the results of our method are more plausible than those of the prior method. (d) The statistics for traffic simulation from an agent's view. Our method has a significantly larger score than method~\cite{chao2018realistic} in the user study with the dataset. Our method also shows better performance in the user study without the dataset. (e) The statistics of the user study for the comparison of VR and desktop. The scores are transferred so that VR is supposed on the left. The scenarios shown through VR have better scores.
}
\label{fig:userStudy}
\end{figure*}

\textbf{Experiment Goals \& Expectations:}
For the first user study, we hypothesize that the results simulated by our method will exhibit more plausible movements than prior multi-agent methods. For the second user study, we hypothesize that our method results in a better user experience than the prior methods. Therefore, participants will significantly prefer our method over the prior methods in these evaluations. In the third user study, we hypothesize that the results shown in VR can produce a better user experience that those shown on a desktop.

\textbf{Comparison Methods:}
For crowd simulation, we compare our method with the method in~\cite{Karamouzas2017implicit} which is a state-of-art physical-based method for crowd simulation. We also use the dataset~\cite{lerner2007crowds} in crowd simulation. For traffic simulation, we compare our method with the method in~\cite{chao2018realistic}, which is a state-of-art data-driven method on traffic simulation. Here we use the dataset~\cite{NGSIM}. All 2D trajectories generated from simulation methods or extracted from datasets are assigned to 3D characters. We also compare mixed traffic results shown in VR and those shown on a desktop.

\textbf{Environments:}
In the first and second user study, we used three scenarios for crowd simulation. The scenario with the dataset~\cite{lerner2007crowds} is in a street with 18 agents. The other two scenarios are the one in which two crowds (100 agents in total) encounter each other and the scenario in which 36 agents are located on a circle moving towards the opposite positions.
We also use three scenarios for traffic simulation. The scenario with the dataset~\cite{NGSIM} is on a straight 4-lane road with 156 agents. The other two scenarios are on a twisting 2-lane road with 80 agents and on a twisting 4-lane road with 200 agents. In the third user study, we use one instance for the scenario with 50 cars and a car's view. We also use three instances for the scenario with 35 cars and 30 pedestrians. In each instance, we use different agent views: one from a car's view, one from the view of a pedestrian walking on a zebra crossing, and one from the view of a pedestrian walking on a sidewalk.

\textbf{Experimental Design:}
We conduct the user studies based on a paired-comparison design. For the scenarios with a dataset, we design two comparison pairs: the dataset vs. our method, and the dataset vs. the prior method. We design one comparison pair for each scenario without a dataset: our method vs. the prior method. For each pair, we show two pre-recorded videos in a side-by-side comparison. The order of the scenarios was random. The position (left or right) of each method was also random. For the scenarios for VR vs desktop comparison, we ask the participants to answer the questionnaire after see the scenarios via VR and the scenarios via desktop.

\textbf{Metrics:}
In each user study, participants were asked to choose a score using a 7-point Likert scale, in which 1 means that the result presented on the left is strongly plausible, 7 means that the result presented on the right is strongly plausible, and 4 means no preference for either method. To combine the user study results in the same scale, we transfer the score for each method to a certain side when we deal with the scores.

\subsection{User Study with an Overhead View}
The user studies for crowd simulation and traffic simulation with an overhead view were completed by 26 participants (15 females and 11 males). We performed two-sample t-tests for the scenarios with datasets (one for crowd simulation and another for traffic simulation). We hypothesize that the mean value of our method is bigger than that of the prior method. Meanwhile, we performed one-sample t-tests for the scenarios without datasets (two scenarios for crowd simulation and two for traffic simulation), hypothesizing that the mean value of our method is bigger than 4, which indicates no difference. Overall, participants believed that our method was more plausible than the compared methods for both crowd simulation and traffic simulation. Fig.~\ref{fig:userStudy} (a)-(b) shows details about the scores for each comparison.

\textbf{User Study for Crowd Simulation}
For the scenario with the dataset, our method's mean score is significantly larger than the prior method's mean plausibility score ($t(25) = 2.9111$, $p=0.0027<0.01$). For the scenarios without datasets, our method's mean score shows a significant difference from the hypothetical mean ($t(51) = -8.7555$, $p<0.001$).

\textbf{User Study for Traffic Simulation}
For the scenarios with datasets, our method's mean of the score is significantly larger than the prior method's mean plausibility score ($t(25) = 2.4422$, $p=0.0091<0.01$). For the scenarios without datasets, our method's mean score shows a significant difference from the hypothetical mean ($t(51) = -3.0169$, $p=0.002<0.01$).

\subsection{User Study with an Agent View}
The user studies for crowd simulation and traffic simulation from an agent's view were completed by 28 participants (17 females and 11 males).
For the user study from an agent view, we also performed two-sample t-tests for the scenarios with datasets hypothesizing that our method has a larger mean score than the prior method. For the scenarios without datasets, we performed one-sample t-tests hypothesizing that the mean value of our method is larger than 4 (no difference). Overall, participants also judged that our method is more plausible than the prior methods. The statistics of the participants' plausibility evaluations can be found in Fig.~\ref{fig:userStudy} (c)-(d).

\textbf{User Study for Crowd Simulation}
For the scenario with a dataset, the mean plausibility score of our Heter-Sim is significantly larger ($t(27) = 2.6692$, $p=0.005<0.01$) than the method~\cite{Karamouzas2017implicit}. The mean score of our method has a significantly superior to the hypothetical mean ($t(55) = -5.0281$, $p<0.001$) for the scenarios without datasets.

\textbf{User Study for Traffic Simulation}
For the scenario with a dataset, the mean score of our method is significantly larger than the mean score of the prior method ($t(27) = 6.4890$, $p<0.001$). For the scenarios without datasets, the mean score of our method shows a significant difference from the hypothetical mean with $t(55) = -8.0381$ and $p<0.001$.

\subsection{User Study via VR or desktop}
The user studies for the comparison between VR and desktop were taken by 28 participants (14 females and 14 males). We performed one-sample t-tests for the four instances by hypothesizing that the mean score of VR is bigger than 4 (no difference). Overall, participants believed that the results shown with VR are more plausible than those shown with a desktop. Fig.~\ref{fig:userStudy} (e) shows the details about the scores for each comparison. In each scenarios, the score of VR is significantly better than that of desktop. $t(27) = -5.0138$, $p<0.001$ for the first scenario, $t(27) = -4.16478$, $p<0.001$ for the second scenario, $t(27) = -3.9890$, $p<0.001$ for the third scenario, and $t(27) = -5.7564$, $p<0.001$ for the last scenario. In total, the mean score for VR shows a significant difference from the hypothetical mean ($t(111) = -9.3485$, $p<0.001$).


\section{Conclusion, Limitation and Future work}
We present a novel and general data-driven optimization method that can generate plausible behaviors for heterogeneous agents in different scenarios. We demonstrate our model's generalizability by simulating human crowds, traffic, and that mixes traffic in multiple scenarios. To the best of our knowledge, this is the first data-driven multi-agent method that is applicable to such different simulation scenarios and mix different kinds of agents (e.g., vehicles and pedestrians).

The simulation results of our method are plausible. We compare our results with prior methods in the same scenarios and by conducting three user studies with various scenarios from different views and analyzing the statistical results of the user studies. Our method can generate results that are closer to the original datasets, than those achieve with the prior methods. In addition, our model is fast and can be used for interactive simulations (Tab.~\ref{tab:runTime}). We also demonstrate that the plausibility of our method can be increased via VR by performing a user study comparing the results via VR or desktop.

Our method can simulate behaviors that are different from those of the input datasets. First, our method can generate larger and denser groups than those in the input datasets (Fig.~\ref{fig:crowd}). Second, our method can simulate scenarios that may differ from those of the input datasets (Fig.~\ref{fig:crowd} (b), Fig.~\ref{fig:traffic} (a)). Third, our method can mix different kinds of agents in the same scenario (Fig.~\ref{fig:traffic} (b) and (c)).

\noindent {\bf Limitations:} Although our approach can generate various behaviors even with a simple, sparse input dataset, the actual performance of our approach can vary based on the datasets. For example, if the dataset only has two magnitudes of velocity in it, the velocity of a car attempting to stop and move again after several seconds will not be continuous.
Because our method uses a forward Euler integration scheme, the stability of our simulation depends on the size of the timestep. An implicit integration scheme~\cite{Karamouzas2017implicit} can be introduced to improve the stability.
We represent agents as rectangles or circles. More precise geometrical shapes should be used to implement better collision avoidance.

As part of future work, our work can be extended in many ways. The input data is not limited to the real datasets and users can also use simulation results to direct certain behaviors. Therefore, the variety or diversity of simulation results can be dramatically increased. We could add traditional context-aware methods to our work to create a variety of behaviors in multiple agents, which would improve the realism of the simulation results. The idea of reducing the solution space according to the continuity of movement can be applied to optimization problems in animation.

Our model can be extended to other areas. The key idea of our method can be extended to data-driven methods to simulate other particle systems, such as fluid simulation~\cite{chu2017data,fu2017polynomial} and cloth simulation. If we treat the vertex as the agent in our system and the connection between vertices as the relationship, our framework can also be applied to data-driven body animation~\cite{kim2017data-driven}. Because we model the decision-making process as an energy-based optimization problem, this idea may be applicable to path planning for robotics and unmanned aerial vehicles. Finally, we want to further evaluate the benefits of our simulator in VR and training scenarios.

\bibliographystyle{abbrv-doi}

\bibliography{dataDriven}
\end{document}